\documentclass[twocolumn,floatfix,aps,superscriptaddress]{revtex4-1}

\usepackage{graphicx}
\usepackage{amsmath}
\usepackage{amssymb}
\usepackage{bm}
\usepackage{color}
\usepackage{comment}
\usepackage[FIGTOPCAP]{subfigure}
\usepackage[]{footmisc}

\begin{document}

\title{Phase shift of cyclotron orbits at type-I and type-II multi-Weyl nodes}

\author{M. Breitkreiz}
\affiliation{Instituut-Lorentz, Universiteit Leiden, P.O. Box 9506, 2300 RA Leiden, The Netherlands}
\affiliation{Dahlem Center for Complex Quantum Systems and Fachbereich Physik, Freie Universit\" at Berlin, 14195 Berlin, Germany}
\author{N. Bovenzi}
\affiliation{Instituut-Lorentz, Universiteit Leiden, P.O. Box 9506, 2300 RA Leiden, The Netherlands}
\author{J. Tworzyd{\l}o}
\affiliation{Institute of Theoretical Physics, Faculty of Physics, University of Warsaw, ul.\ Pasteura 5, 02--093 Warszawa, Poland}

\date{March 2018}

\begin{abstract}
Quantum oscillations of response functions in high magnetic fields tend to  
reveal some of the most interesting properties of metals. In particular,
the oscillation phase shift is sensitive to topological band features, 
thereby helping to identify the presence of Weyl fermions. 
In this work we predict characteristic parameter dependence of the phase shift for Weyl fermions 
with tilted and overtilted dispersion 
(type I and type II Weyl fermions) and an arbitrary topological charge (multi-Weyl fermions). 
For type-II Weyl fermions our calculations capture the case of magnetic breakthrough 
between the electron and the hole part of the dispersion. Here the phase shift 
turns out to depend only on the quantized topological charge due to cancellation of non-universal contributions from the 
electron and the hole part.
\end{abstract}
\maketitle

\textit{Introduction}---Electrons moving along cyclotron orbits in a homogeneous magnetic field 
are subject to the quantization condition \cite{Roth1966}
\begin{equation}
l^2 S= 2\pi(m+\gamma),\;\;\;\; m\in  \mathbb{Z},
\label{qc}
\end{equation}
where $S$ is the zero-field area enclosed
by the cyclotron orbit in momentum space, $l=\sqrt{\hbar/eB}$
is the magnetic length, and the offset $\gamma $ includes quantum corrections, which  
can be expanded in powers of the magnetic field $B$ \cite{Gao2017}. 
In the semiclassical regime when the magnetic
length is much larger than the Fermi wavelength, field-dependent corrections to
$\gamma$ are suppressed and the remaining number of zeroth order in $B$ 
encodes valuable information about the electronic properties of the system.
In particular, the offset includes contributions coming from topological features in the band structure
\cite{Mikitik1998, Mikitik1999, Fuchs2010, Alexandradinata2018}, 
which makes it the subject of high current interest.
Experimentally it can be deduced from  quantum oscillations in the de Haas-van 
Alphen or the Shubnikov-de Haas effects, widely used nowadays to identify 
Weyl, Dirac, and nodal-line semimetals  \cite{Zhang2005, He2014, Huang2015a, Hu2016, Pezzini2017}.

Interestingly, in some well-studied systems the offset measures the topological features independent of the specifics 
of the band structure. So, e.g., in  graphene and graphene bilayer exposed to an out-of-plane magnetic field
the offset turns out to be given by a winding 
number---the number of full turns made by the direction of the electron's pseudospin degree of freedom 
during a single turn around the cyclotron orbit \cite{Mikitik1998, Fuchs2010}. This integer winding number is a robust feature, 
determined by the type of the band touching, and is sometimes called the topological charge of the 
Weyl or Dirac fermion \cite{Burkov2016}.
In contrast to the common belief, however, 
 the topological charge contributes to the offset in such a robust manner only in exceptional cases,
namely when particular symmetry constraints are satisfied \cite{Alexandradinata2018}. 
 In general, the offset is sensitive also to other parameters of the band touching and it is the aim of this
work to characterize this sensitivity.

\begin{figure}[b]
\includegraphics[width=0.7\columnwidth]{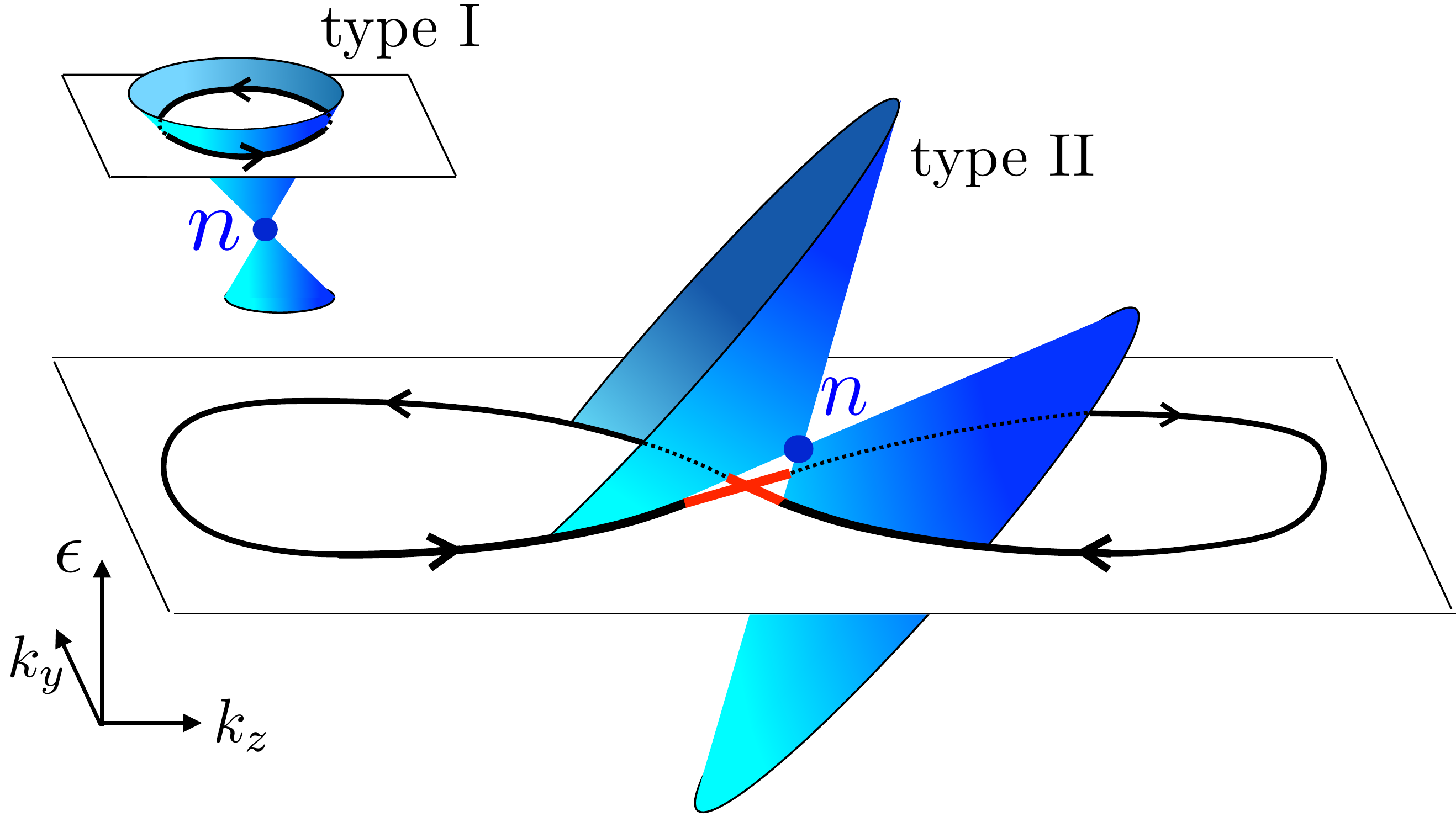}
\caption{Schematic illustration of a breakthrough cyclotron orbit (figure-8 curve) at a type-II Weyl node
with topological charge $n$. The red part indicates quantum tunneling in the magnetic-breakthrough region.
The inset shows a cyclotron orbit at a type-I Weyl node.}
\label{fig1}
\end{figure}

One important parameter is a linear tilt of the dispersion at the Weyl node, which is generically 
present in material realizations and, most importantly, leads to the occurrence of two types of Weyl nodes, 
as sketched in Fig.\ \ref{fig1}. Upon the type-I to type-II transition, the tilt exceeds a critical value, above which
an equi-energy surface near the node cuts both bands \cite{Soluyanov2015}. The closed cyclotron orbit 
at a type-I Weyl node is thereby replaced by two open branches, which can be closed at large momenta by higher-order 
corrections to the Weyl Hamiltonian, resulting in two cyclotron orbits, one electron-like and one hole-like. 
Band details determine a critical magnetic field, above which
the two separate cyclotron orbits effectively merge into a single orbit via magnetic breakthrough
\cite{Cohen1961, Kaganov1983}. 
This critical field is zero if the energy and the 
parallel momentum are exactly at the node 
where the two contours touch 
\cite{OBrien2016, Alexandradinata2017a}, and is larger than zero if the gap between the contours is finite. 
The magnetic breakdown contributes an additional phase to the offset $\gamma$, so one would expect that the offset 
is even more sensitive to details of the orbit than in the case without magnetic breakdown.

In this work we analyze the offset for orbits at both types of Weyl nodes 
and find characteristic dependence of $\gamma$ on the Weyl-node parameters. Most surprisingly,
the offset of the breakthrough orbit at a 
type-II Weyl point turns out to depend only on the topological charge. This striking result is based  
on two facts, the universality of the phase jump of $\pi$ acquired in the magnetic-breakthrough region and 
a robust phase shift of $n\pi$ induced by the topological charge. The insensitivity of the latter on details of the 
orbit comes from a cancellation of a non-universal part of the phase in the two loops of the breakthrough orbit, 
which are traversed in opposite directions. 

\textit{Model}---We consider a set of Hamiltonians that govern the  physics close to topologically distinct 
band touchings,
\begin{subequations}
\begin{align}
H_0 ={}& k_-\sigma_++k_+\sigma_- + u\, k_z\sigma_0 \label{H0}, \\
H_n ={}& k_-^n\sigma_++k_+^n\sigma_- + u\, k_z\sigma_0 +k_z\sigma_z,\; n \in \{1,2,\dots\}, \label{Hn}
\end{align}
\end{subequations}
where $k_\pm=k_x\pm i k_y$, $(k_x,k_y,k_z)=\mathbf{k}$ are momenta (scaled by velocities),
$\sigma_\pm=\sigma_x\pm i\sigma_y$, $\sigma_{x,y,z}$ are Pauli matrices, and
$\sigma_0$ the identity matrix. 
The band touching at $\mathbf{k}=0$ described by $H_n$ corresponds to a topologically 
protected multi-Weyl node of order $n$ \cite{Fang2012}, while  
$H_0$ describes a trivial, non-protected band touching (a gap is produced by a perturbation $\propto \sigma_z$). 
The parameter $u>0$ controls
the tilt of the Weyl cone; for $u<1$ and  $u>1$ the Weyl cone is of type I and II, respectively. 

The magnetic field pointing in $x$ direction moves the particles along equi-energy contours
$k_z(k_y)$ at fixed energy $\epsilon$ and parallel momentum component $k_x$.
The contours are determined by the Schr\" odinger equation
\begin{equation}
H_n|u_{n\pm}\rangle=\epsilon |u_{n\pm}\rangle,
\label{schroedinger}
\end{equation}
where $\pm$ denote the two bands.

In the quantization condition \eqref{qc} one can distinguish three phase shifts that contribute to the offset
\begin{equation}
\gamma=\frac{1}{2\pi}(\phi_0+\phi_\mathrm{b}+\phi_\mathrm{t}).
\label{gamma}
\end{equation}
Here $\phi_0$ and $\phi_\mathrm{b}$ are phase shifts that occur at singular points on the orbit. Specifically,
turning points give rise to the Maslov phase $\phi_0$ 
\cite{Keller1958}, in which each turning point contributes 
a phase jump of $\pm\pi/2$, the sign determined by the sign of the curvature at the turning point. 
In particular one finds that $\phi_0=\pi$ and $\phi_0=0$ for orbits that can be deformed into a 
circle and into an 8-shape, respectively.
With $\phi_\mathrm{b}$ we denote the phase shifts that occur due to magnetic breakdown. 
Finally,  $\phi_\mathrm{t}$ is the topological phase shift, which includes the Berry phase accumulated during a full turn around
the orbit and the effect of the orbital magnetic moment 
\cite{Mikitik1999, Panati2003, Fuchs2010, Alexandradinata2018}. The explicit calculation of   
$\phi_\mathrm{b}$ and $\phi_\mathrm{t}$ is the main result of this work, which will be presented in
the following.

\textit{Topological phase shift}---The topological phase shift of a closed contour at energy
$\epsilon$ and the fixed momentum component $k_x$
is given by \cite{Fuchs2010, Xiao2010}
\begin{equation}
\phi_\mathrm{t} =\oint dk_y'\bigg[A-\frac{d k_z(k_y')}{d\epsilon}M\bigg].\label{phiB}
\end{equation}
Here the first term is determined by the Berry connection
projected onto the contour,
\begin{align}
A =& i\langle u|\nabla_\mathbf{k}|u\rangle\cdot \frac{d\mathbf{k}}{dk_y}
 = i\langle u|\frac{d}{dk_y}|u\rangle,
 \label{A}
\end{align}
which contributes to $\phi_\mathrm{t}$ the usual Berry phase of the closed orbit. 
The second term is the correction to the zero-field area
$S$ coming from the orbital magnetic moment projected onto
the direction of the magnetic field \cite{Panati2003},
\begin{align}
M={}& \frac{i}{2} \Big[ \big(\partial_{k_y} \langle u|\big)\big(\epsilon-H\big) \big(\partial_{k_z}|u\rangle\big) 
\nonumber \\
 & -\big(\partial_{k_z} \langle u|\big)\big(\epsilon-H\big) \big(\partial_{k_y}|u\rangle\big)
 \Big].
 \label{M}
\end{align}

The eigenfunctions of the Hamiltonian $H_i$ can be written as
\begin{align}
|u_{0\,\pm}\rangle={}& \frac{1}{\sqrt{2}}\begin{pmatrix}
\mp e^{-i\alpha} \\ 1\end{pmatrix}, \nonumber\\
|u_{n+}\rangle={}& \begin{pmatrix}
-\sin\tfrac{\beta}{2}e^{-i n \alpha} \\ \cos\tfrac{\beta}{2} \end{pmatrix},\;\;
|u_{n-}\rangle= \begin{pmatrix}
\cos\tfrac{\beta}{2}e^{-i n\alpha} \\ \sin\tfrac{\beta}{2} \end{pmatrix},
\end{align}
where the angles $\alpha$ and $\beta$ are defined as
\begin{align}
\cos\beta={}& \frac{k_z}{k},\;\; \sin\beta=\frac{\big(k_x^2+k_y^2\big)^{\tfrac{n}{2}}}{k}, \nonumber\\
\alpha ={}& \mathrm{Arg}(k_x+i k_y),\;\;
k=\sqrt{\big(k_x^2+k_y^2\big)^n+k_z^2}.
\label{us}
\end{align}
For the topologically trivial case we obtain from \eqref{A}--\eqref{us} 
\begin{align}
A_{0\pm}={}&\frac{k_x}{2(k_x^2+k_y^2)},\;\;\;\;M_{0\pm}=0,
\end{align}
and the topological phase shift vanishes as it should,
\begin{equation}
\phi_\mathrm{t}^{\pm}=\oint dk_y'\,A_{0\pm}=\oint dk_y'\frac{k_x}{2[k_x^2+(k_y')^2]}=0\;\;\;\;(n=0),
\end{equation}
independent of the integration contour.
For the non-trivial case, we obtain
\begin{align}
A_{n\pm} ={}& \frac{nk_x\big(k_x^2+k_y^2\big)^{n-1}}
{2k(k\pm k_z)}, \,\; 
M_{n\pm}= -\frac{nk_x\big(k_x^2+k_y^2\big)^{n-1}}{2k^2}.
\label{Mn} 
\end{align}
To calculate the topological phase shift, we consider the explicit expression for the equi-energy contours, which is derived from \eqref{schroedinger}
in the form
\begin{align}
k_z^{\pm}(k_y)={}& \frac{\epsilon\, u\pm\sqrt{(u^2-1)\big(k_x^2+k_y^2\big)^n+\epsilon^2}}{u^2-1}.
\label{cont}
\end{align}
For $u>1$, the contours given by $k_z^{\pm}(k_y)$ are disjoint and we need to introduce 
an additional orbit segment that connects the two open ends of 
$k_z^{\pm}(k_y)$ at $k_z\to\pm\infty$. 
These connecting segment can be realized by 
an additional mass term $\eta k_z^3\sigma_z$ in the 
Hamiltonian, with an infinitesimal $\eta>0$. The reconnection 
then occurs at large momenta $k_z$, with $|k_z|>(u-1)/\eta\to\infty$. In the expressions  \eqref{Mn}
for $A$ and $M$ the additional mass term replaces
$k_z\to k_z+\eta k_z^3$. On the connecting segment,
$A$ and $M$ go to zero like $\eta^2$, while
the integration along the connecting segment 
gives a factor of order $1/\eta$. 
Hence the contribution of the connecting segment to $\phi_\mathrm{t}$
vanishes and the integration reduces to the
integration along the main contour $k_z^{\pm}(k_y)$. 

Inserting \eqref{Mn} and \eqref{cont} into
\eqref{phiB} we obtain
\begin{align}
\phi_\mathrm{t}^{\pm}=\mp\int dk_y'
\frac{(u+1)nk_x\big(k_x^2+(k_y')^2\big)^{n-1}}
{2\big[k^{\pm}\pm k_z^{\pm}\big]\big[k_z^{\pm}\mp uk^{\pm}\big]}.
\label{phib}
\end{align}
For a type-II cone ($u>1$) we use the substitution $\kappa=k_y'/k_x$ and obtain
\begin{align}
\phi_\mathrm{t}^{\pm} ={}& \int_{-\infty}^\infty d\kappa
\frac{n(\kappa^2+1)^{n-1}}{2\sqrt{(\kappa^2+1)^n+\cot^2\theta}}\nonumber\\
&{}\times \Big(\sqrt{(\kappa^2+1)^n+\cot^2\theta}\pm\cot\theta\Big)^{-1},
\label{ia}
\end{align}
where the parameter $\theta$ encoding contour details is defined as
\begin{equation}
\theta=\begin{cases}
\mathrm{atan}\bigg(\frac{k_x^n\sqrt{u^2-1}}{\epsilon}\bigg) 
& u>1\\
\mathrm{atanh}\bigg(\frac{k_x^n\sqrt{1-u^2}}{\epsilon}\bigg) & u<1. \end{cases}
\label{theta}
\end{equation}
The integral in \eqref{ia} needs to be calculated numerically (see below); for the special case $n=1$,
 we find the closed-form solution
\begin{align}
\phi_\mathrm{t}^{\pm}=\frac{\pi}{2}(1\mp\mathrm{sign}\theta)\pm\theta \;\;\;\;\;\;(n=1).
\label{phibn1}
\end{align}

While $\phi_\mathrm{t}^{\pm}$ are the topological phase shifts of the two 
(electron/hole) orbits $k_z^\pm(k_y)$,  
the sum $\phi_\mathrm{t}^{+}+\phi_\mathrm{t}^{-}\equiv \phi_\mathrm{t}^\mathrm{br}$ is the topological phase shift of the 
breakthrough orbit, i.e., the figure-of-8 orbit that encloses both the electron and the hole pocket. 
Using the substitution 
$z=(\kappa^2+1)^n$ the integral for $\phi_\mathrm{t}^\mathrm{br}$ simplifies to
\begin{equation}
\phi_\mathrm{t}^\mathrm{br}=\int_1^\infty dz\frac{1}{\sqrt{z^{\frac{2n+1}{n}}-z^2}}=n\pi,
\label{phisum}
\end{equation} 
where the $\theta$ dependent part cancels out. As a result the topological phase shift of the figure-of-8 
orbit only depends on the quantized topological charge $n$, in contrast to the $\theta$-dependent phase shifts of 
the separate orbits. 

For type-I Weyl fermions ($u<1$) $k_z^\pm$ are two parts of a single
closed contour, which topological phase is denoted $\phi_\mathrm{t}$.
A closed-form solution for the integral \eqref{phib} is found  
for $n=1$,
\begin{equation}
\phi_\mathrm{t}=\pi\,\mathrm{sign}\theta\;\;\;\;\;(n=1),
\end{equation}
in agreement with Refs.\ \cite{Mikitik1998, Fuchs2010, Yu2016, *Tchoumakov2016, *Udagawa2016, *Mikitik2016}.
For $n\ge 2$, we find in the limits $\theta\to 0^\pm $ and $\theta\to\pm\infty$,
\begin{equation}
\phi_\mathrm{t}=\xrightarrow{\theta\to 0^\pm}
n\,\pi\,\mathrm{sign}\theta,\;\;\;
\phi_\mathrm{t}=\xrightarrow{\theta\to \pm\infty}
\sqrt{n}\,\pi\,\mathrm{sign}\theta.
\end{equation}
The  full $\theta$ dependence will be discussed below. 

\textit{Breakthrough phase shift}---To calculate the additional phase 
shift of the figure-of-8 orbit due to magnetic breakdown, we follow a standard route \cite{Kaganov1983} 
and calculate the scattering matrix that relates the exact wavefunction of the 
magnetic-breakdown region with the in- and out-going semiclassical wavefunctions. 

We start with the non-topological Hamiltonian $H_0$.
Introducing the magnetic field via Peierls substitution 
$k_z\mapsto k_z+il^{-2} \partial_{k_y}$, followed by a 
unitary transformation,
\begin{equation}
\tilde{H}_0 = e^{-il^2 (k_z-\epsilon/u) k_y}H_0 e^{il^2 (k_z-\epsilon/u) k_y},
\end{equation}
we arrive at
\begin{equation}
\tilde{H}_0 = k_x\sigma_x+k_y\sigma_y + i\,u\, l^{-2} \partial_{k_y}\sigma_0+\epsilon.
\end{equation}

Rescaling the variables as $k=l k_y/\sqrt{u}$, $\delta_0=l k_x/\sqrt{u}$,  the 
 Schr\" odinger equation $\tilde{H}_0\psi=\epsilon\psi$  reads
\begin{equation}
\big[\sigma_x\delta_0+ \sigma_y k +i  \partial_{k}\big]\psi=0.
\label{hamful}
\end{equation}
The exact solution of  \eqref{hamful} is known from the Landau-Zener problem 
\cite{Landau1977}. To obtain the phase shift in comparison to the semiclassical solution
of \eqref{hamful}, the
exact wavefunctions are matched with the incoming semiclassical wavefunctions at 
$k\ll-\delta_0$, denoted $\psi_i^\pm$, and outgoing  $\psi_f^\pm$ at $k\gg \delta_0$.
From this standard procedure (recapitulated in the Supplemental Material) 
we obtain the scattering matrix $S$ that relates the final state in the basis $(\psi_f^+,\psi_f^-)$ 
to the incoming state in the basis $(\psi_i^+,\psi_i^-)$, 
\begin{align}
S={}& \begin{pmatrix}
\sqrt{1-W}\,e^{i\alpha} & -i\sqrt{W} \\
-i \sqrt{W} &  \sqrt{1-W}\,e^{-i\alpha} \end{pmatrix}, 
\label{scatmattf}
\end{align}
where 
\begin{equation}
W= e^{-\pi\delta_0^2 },\;\;
 \alpha= \tfrac{\pi}{4}+\tfrac{\delta_0^2}{2}-\tfrac{\delta_0^2}{2}\ln\tfrac{\delta_0^2}{2}
+\mathrm{arg}\,\Gamma \Big(i\tfrac{\delta_0^2}{2} \Big).
\end{equation}
The breakthrough orbit dominates if $\delta_0\ll 1$, $W\approx1$, in which case each band transition in the breakthrough  
region contributes  a phase jump of $\pi/2$ giving in total the phase shift $\phi_\mathrm{b}=\pi$ for the 
breakthrough orbit.

 \begin{figure}[t]
\includegraphics[width=0.64 \columnwidth]{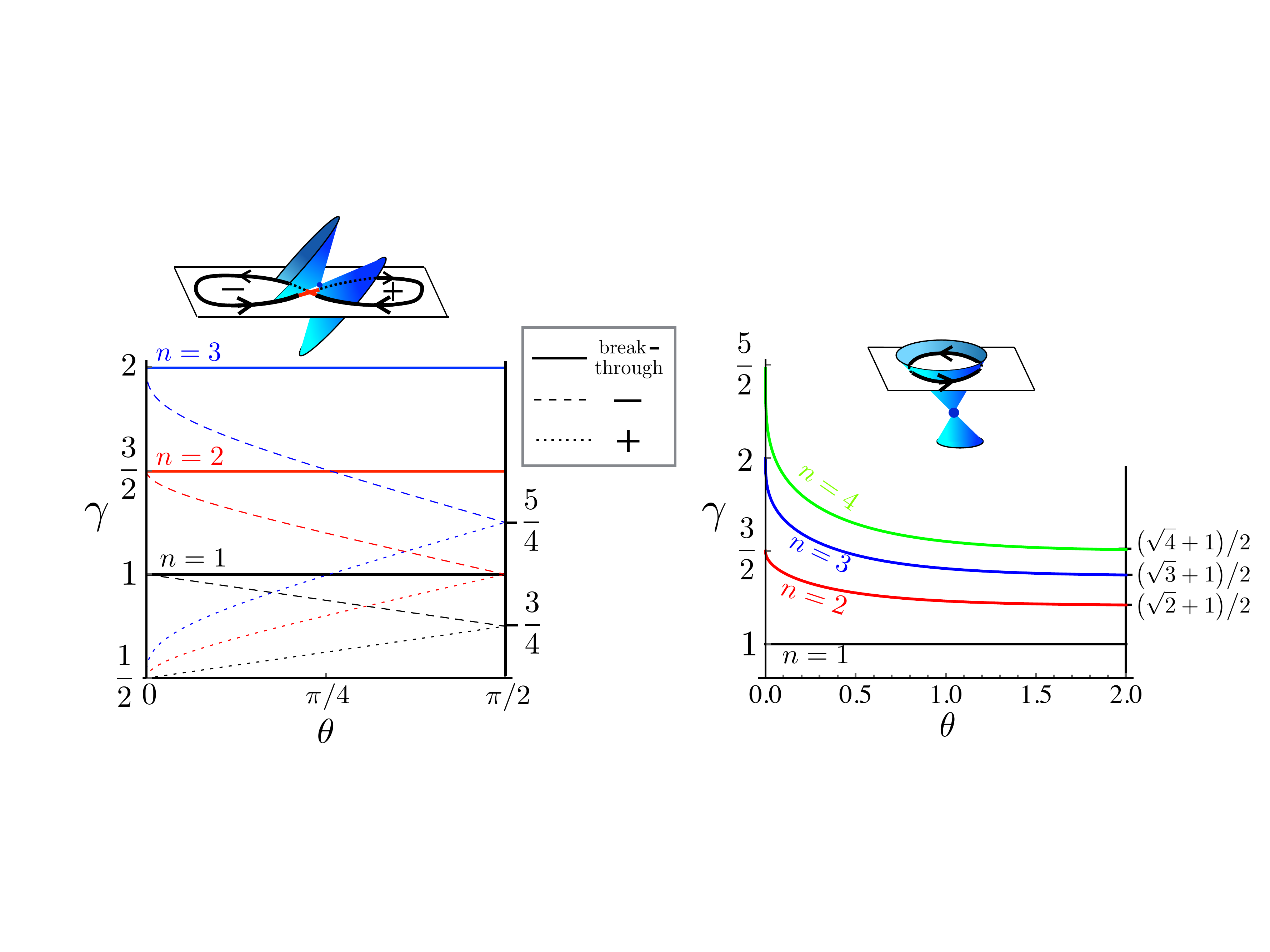}
\caption{Parameter dependence of the offset $\gamma$ of orbits at a type-I Weyl node.}
\label{phibI}
\end{figure}

 \begin{figure}[h]
\includegraphics[width=.75 \columnwidth]{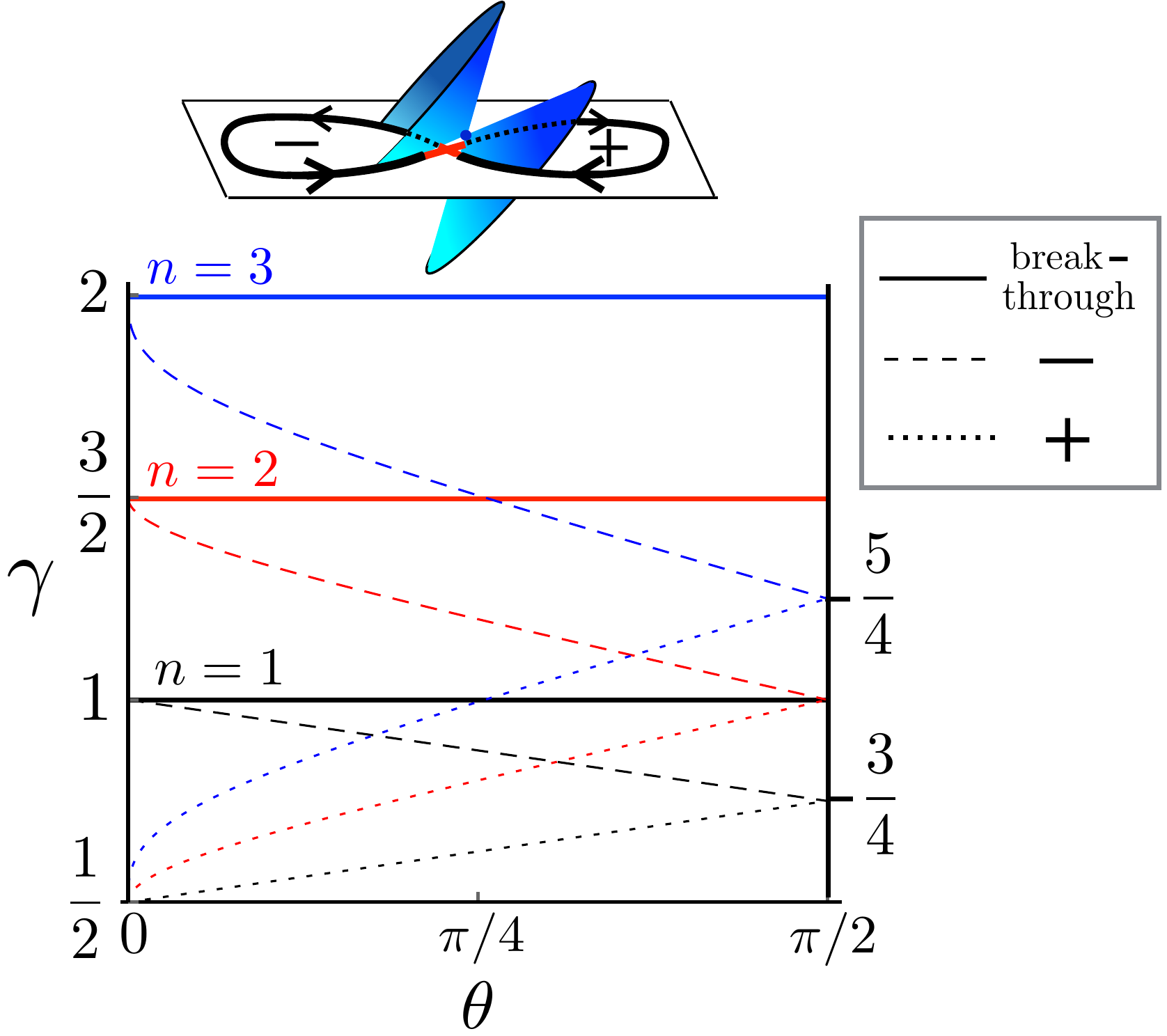}
\caption{Parameter dependence of the offset $\gamma$ of orbits at a type-II Weyl node. 
The offsets of separate orbits $k_z^+$ and 
$k_z^-$ (without magnetic breakthrough) depend on the band parameter $\theta$, while the offset
 of the figure-8 breakthrough orbit only depends on the topological charge $n$.}
\label{phibII}
\end{figure}

For the topological case, we linearize the Hamiltonian $H_n$ in $k_y$, leading to
\begin{equation}
H_n'=k_x^n\sigma_x+nk_x^{n-1}k_y\sigma_y+k_z\sigma_z+u k_z.
\label{hnp}
\end{equation}
After Peierls substitution we apply the
unitary transformation given by 
\begin{equation}
\tilde{H}_n = e^{-il^2  [k_z-\epsilon/(u^2-1)] k_y}H_n' e^{il^2 [k_z-\epsilon/(u^2-1)] k_y}.
\label{kzshift}
\end{equation}
Rescaling and transforming the variables as
\begin{subequations}
\begin{align}
k={}& l k_y (u^2-1)^{-1/4}\sqrt{nk_x^{n-1}}, \\
\delta_n={}& l\,\mathrm{sign}(\epsilon) \frac{\sqrt{\epsilon^2+(u^2-1)k_x^{2n}}}{(u^2-1)^{3/4}k_x^{(n-1)/2}},
\label{delta_n}
\end{align}
\end{subequations}
we obtain the Schr\" odinger equation
\begin{align}
& \Big[\delta_n \sqrt{ u^2-1} \sin\theta \sigma_x+
 k \sqrt{ u^2-1}\sigma_y \nonumber\\
&\bigskip + i\partial_k (u  + \sigma_z) + 
 \delta_n \cos\theta (1 + u \sigma_z)\Big]\psi=0.
\label{seII}
\end{align}
Multiplying \eqref{seII} from the left with ${\cal M}=\mathrm{diag}[(u+1)^{-1},(u-1)^{-1}]$ and applying
a transformation given by 
\begin{equation}
T=-i\begin{pmatrix}
\frac{1-u}{\sqrt{u^2-1}} && \frac{u-1}{\sqrt{u^2-1}} \\ 1 && 1 
\end{pmatrix} \sigma_z e^{-i \sigma_y \theta/2},
\end{equation}
we again arrive at the differential equation of the Landau-Zener form \eqref{hamful},
\begin{align}
\hat{H} \hat{\psi}(k) &=\big(\delta_n\, \sigma_x+ k\, \sigma_y + i\partial_k\big)\hat{\psi}(k)= 0,
\end{align}
where $\hat{\psi}(k) = T^{-1}\psi(k)$ and $\hat{H} = T^{-1}\, {\cal M}\,\tilde{H} T$.
The solution of \eqref{seII} is thus given by the solution of the Landau-Zener problem multiplied from the
left with the matrix $T$. Note that the $\theta$ phase brought into the full solution by the matrix $T$
is the topological phase of the full solution induced by the non-trivial topology of the Hamiltonian. 

The $S$ matrix is obtained by matching $\psi(k)$ with the semiclassical
solution of \eqref{seII}. Since $H_n'$ is topologically equivalent 
to $H_1$ (note that the dynamical variables are $k_y$ and  $k_z$, while $k_x$ is fixed), the topological 
phase shift of the semiclassical solution is given by \eqref{phibn1}, which cancels the $\theta$ phase of the
full solution and the result is the same $\theta$-independent scattering matrix \eqref{scatmattf}, 
with $\delta_0$ replaced by $\delta_n$. In particular, the breakthrough phase shift $\phi_\mathrm{b}=\pi$ 
 also holds in the topological case.

\textit{Discussion}---Having thus calculated  the phase shifts, we now show the full $\theta$-dependence of
the offset $\gamma$, defined in Eq.\ \eqref{gamma}, in Figs.\ \ref{phibI} and \ref{phibII}
 \cite{*[{In practice, the offset can only be measured modulo one, corresponding to one Landau-level spacing. 
Nevertheless in Figs.\ \ref{phibI} and \ref{phibII} we plot the full $\gamma$ for clarity of the graphic}] [{}] dummy2}. 
For the figure-of-8 orbit, the magnetic breakthrough contributes an
offset $1/2$ and the topological charge adds an extra contribution $n/2$. The $\theta$ independence 
is based on the cancellation of the $\theta$-dependent parts from the hole and the electron pockets.
The universality of the breakthrough phase shift is, instead, less surprising, since the same universal value was found previously for 
non-topological band touchings \cite{Kaganov1983}. 

In contrast, without breakthrough (dashed/dotted curves in Fig.\ \ref{phibII}) or in case of a type-I Weyl node
(Fig.\ \ref{phibI}), the offset has a non-trivial
dependence on the orbit details that are encoded in $\theta$. The only exception is the   
case $n=1$ of the type-I Weyl node, which shows no $\theta$ dependence owing to the higher 
symmetry of the dispersion \cite{Alexandradinata2018}. This is also the only case with a known full quantum-mechanical
solution \cite{Mikitik1998, Fuchs2010, Yu2016, *Tchoumakov2016, *Udagawa2016, *Mikitik2016}; it agrees with our semiclassical result. 
In  quantum oscillation experiments, the measured phase shift would likely be averaged over a range of values of the energy and of the  parallel momentum $k_x$, corresponding to a weighted (depending on details of the experimental realization)  average over the parameter $\theta$. In general, this averaging does not destroy the $\theta$ dependence, still allowing to discriminate the 
two cases of quantized and continuously varying $\gamma$.

With regard to the figure-8 breakthrough orbits, our calculations explain recent numerical findings for the offset 
of a thin-film Weyl semimetal  \cite{Bovenzi2017a} and a type-II Weyl semimetal \cite{OBrien2016},
showing, respectively, $\gamma=1/2$ and $\gamma=0$. 
In the case of the thin film, the Hamiltonian at the figure-8 crossing, given in the appendix of Ref.\ \cite{Bovenzi2017a}, 
is equivalent to the non-topological Hamiltonian $H_0$, thus the only phase contributing is the breakthrough phase $\phi_\mathrm{b}=\pi$,
which explains the offset $\gamma=\phi_\mathrm{b}/2\pi=1/2$. In case of the  type-II Weyl semimetal, the Hamiltonian 
is equivalent to $H_1$, where the additional topological phase $\phi_\mathrm{t}=\pi$ cancels the breakthrough phase, which
explains the vanishing offset. This contradicts a previous interpretation that relates the vanishing offset of the latter to a
vanishing Berry phase and neglects the contribution of the breakthrough phase \cite{Alexandradinata2017a}.
In the Supplemental Material 
we present extensions of the numerical calculations to 
the cases $n=2$ and $n=3$, tilted type-I Weyl cones, and several values of $\theta$. 
Also these calculations are in agreement with the analytical results of this work.

\textit{Acknowledgements}---The authors would like to thank C.W.J.\ Beenakker, A.\ Alexandradinata, and M.J.\ Pacholski for valuable discussions. 
This research was supported by the Netherlands Organization for
Scientific Research (NWO/OCW) and an ERC Synergy Grant.

\onecolumngrid

\bibliography{library}

\clearpage

\renewcommand{\theequation}{S\arabic{equation}}
\renewcommand{\thefigure}{S\arabic{figure}}

\setcounter{equation}{0}
\setcounter{figure}{0}

\section*{Supplemental Material}

\section{Topological phase for $u>1$}

While the sum of the phases $\phi_\mathrm{t}^{\pm}$ has been calculated in the 
main text, to obtain each of the phases separately we now focus on the difference.
From \eqref{ia} we find
\begin{equation}
\phi_\mathrm{t}^{-}-\phi_\mathrm{t}^{+}=\int_{-\infty}^\infty d\kappa
\frac{n\cot\theta}{(\kappa^2+1)\sqrt{(\kappa^2+1)^n+\cot^2\theta}}.
\label{a1}
\end{equation}
Using the series expansion
\begin{equation}
\frac{1}{\sqrt{1+q}}=\sum_{m=0}^\infty
\begin{pmatrix}
m-\tfrac{1}{2} \\m \end{pmatrix} (-q)^m
\end{equation}
and the integral
\begin{equation}
\int_{-\infty}^\infty d\kappa\frac{1}{(\kappa^2+1)^\alpha}
=\frac{\sqrt{\pi}\,\Gamma\big(\alpha-\tfrac{1}{2}\big)}
{\Gamma(\alpha)},
\end{equation}
Eq.\ \eqref{a1} can be written as
\begin{align}
\phi_\mathrm{t}^{-}-\phi_\mathrm{t}^{+}
 ={}& n\cot\theta
\sum_{m=0}^\infty (-1)^m
\begin{pmatrix} m-\tfrac{1}{2} \\m \end{pmatrix} \nonumber\\
&{}\times\frac{\sqrt{\pi}\,\Gamma\big(n m+\tfrac{n+1}{2}\big)}
{\Gamma\big(n m+\tfrac{n+2}{2}\big)}(\cot\theta)^{2m}.
\end{align}
For $n=1$ the series is the expansion of $2\arctan(\cot\theta)/\cot\theta$, which gives
\begin{equation}
\phi_\mathrm{t}^{-}-\phi_\mathrm{t}^{+}= \mathrm{sign}(\theta)\pi-2\theta.
\end{equation}
Together with \eqref{phisum}, $\phi_\mathrm{t}^{-}+\phi_\mathrm{t}^{+}=n\pi$, this leads to
\begin{align}
\phi_\mathrm{t}^{\pm}=\frac{\pi}{2}(1\mp\mathrm{sign}\theta)\pm\theta.
\label{phibn2}
\end{align}

\section{Topological phase for $u<1$}

For $u<1$, $k_z^\pm$ are two parts of a single
closed contour. The phase $\phi_\mathrm{t}$ of the contour is thus given by the difference
$\phi_\mathrm{t}=\phi_\mathrm{t}^{+}-\phi_\mathrm{t}^{-}$, where, using \eqref{phib}, \eqref{theta}, and 
the substitution $\kappa=k_y'/k_x$,  $\phi_\mathrm{t}^{\pm}$ are given by
\begin{align}
\phi_\mathrm{t}^{\pm} ={}& -\int_{-\kappa_0}^{\kappa_0} d\kappa
\frac{n(\kappa^2+1)^{n-1}}{2\sqrt{\mathrm{coth}^2\theta-(\kappa^2+1)^n}}
\Big(\sqrt{\mathrm{coth}^2\theta-(\kappa^2+1)^n}\pm\mathrm{coth}\theta\Big)^{-1},
\end{align}
where $\kappa_0=\sqrt{(\mathrm{coth}\theta)^{2/n}-1}$. The difference reduces to
\begin{equation}
\phi_\mathrm{t}=
\int_{-\kappa_0}^{\kappa_0} d\kappa
\frac{n\,\mathrm{coth}\theta}{(\kappa^2+1)\sqrt{\mathrm{coth}^2\theta-(\kappa^2+1)^n}}
\end{equation}
and, after the substitution $z=(\kappa^2+1)^n$, can be rewritten as
\begin{equation}
\phi_\mathrm{t}=\int_1^{\mathrm{coth}^2\theta}dz
\frac{\coth\theta}{z\sqrt{\coth^2\theta-z}\sqrt{z^{1/n}-1}}.
\end{equation}
A closed-form solution is found for $n=1$,
\begin{equation}
\phi_\mathrm{t}=\pi\,\mathrm{sign}\theta.
\end{equation}
For a general $n$ we find in the limits $\theta\to 0^\pm $
\begin{equation}
\phi_\mathrm{t} \xrightarrow{\theta\to 0^\pm}
n\,\pi\,\mathrm{sign}\theta
\end{equation}
and $\theta\to\pm\infty$
\begin{equation}
\phi_\mathrm{t}\xrightarrow{\theta\to \pm\infty}
\sqrt{n}\,\pi\,\mathrm{sign}\theta.
\end{equation}

\section{Scattering matrix for magnetic breakdown}

\subsubsection{Non-topological Hamiltonian}

To obtain the full solution in the magnetic-breakthrough region, we solve 
the differential equation
\begin{equation}
\big[\sigma_x\delta+ \sigma_y k +i  \partial_{k}\big]\psi=0.
\label{hamfulA}
\end{equation}
Multiplying from the left with $U=\exp(-i\sigma_x\pi/4)$
and inserting the ansatz  $\psi=U^\dagger(\eta,\xi)^T$ we obtain
\begin{align}
&\big(k^2+\partial_k^2 - i+\delta^2 \big) \eta= 0, \label{diff}\\
 &\xi = -\delta^{-1}\big( k+i\partial_k\big)\eta. \label{xi}
\end{align}
Equation \eqref{diff} can be transformed to Weber's equation for the parabolic cylinder function,
\begin{align}
&\eta''-\big(\tfrac{1}{4}z^2+a\big)\eta=0, 
\end{align}
where
\begin{equation}
z=\sqrt{2}e^{i\pi/4}\,k,\;\;  a=\tfrac{1}{2}+i\gamma,\;\; \gamma=\tfrac{1}{2}\delta^2.
\end{equation}
The two solutions read
\begin{subequations}
\begin{align}
\eta_a ={}& e^{-z^2/4}\; _1F_1\big(\tfrac{1}{2}a+\tfrac{1}{4};\tfrac{1}{2}; \tfrac{1}{2}z^2),\\
\eta_b ={}& z\, e^{-z^2/4}\; _1F_1\big(\tfrac{1}{2}a+\tfrac{3}{4};\tfrac{3}{2}; \tfrac{1}{2}z^2),
\end{align}
\label{phis}
\end{subequations} 
where $_1F_1()$ is the  confluent hypergeometric function. Its general asymptotic form for a large last argument reads
\begin{align}
 & _1F_1(\alpha,\beta, ik^2)
 \xrightarrow{k\to\infty} \Gamma(\beta)\Big(\tfrac{1}{\Gamma(\alpha)}e^{ik^2}(ik^2)^{\alpha-\beta}
 +\tfrac{1}{\Gamma(\beta-\alpha)}(-i k^2)^{-\alpha}\Big).
\end{align}
From this we obtain the asymptotic form of the two solutions
\eqref{phis},
\begin{subequations}
\begin{align}
\eta_a={}& \frac{\Gamma\big(\tfrac{1}{2}\big)}
{\Gamma\big(\gamma\tfrac{i}{2}+\tfrac{1}{2}\big)}
 e^{ik^2/2+i\gamma\ln |k|-\pi\gamma/4}, \\
 \eta_b={}& \mathrm{sign}(k)\,\frac{\sqrt{2}\Gamma\big(\tfrac{3}{2}\big)}
{\Gamma\big(\gamma\tfrac{i}{2}+1\big)}
 e^{ik^2/2+i\gamma\ln |k|-\pi\gamma/4}.
\end{align}
\label{phisol}
\end{subequations}
Inserting into \eqref{xi}, we obtain the two corresponding
expressions for $\xi$,
\begin{subequations}
\begin{align}
\xi_a={}& -\mathrm{sign}(k)\,\frac{\sqrt{2\pi/\gamma}}
{\Gamma\big(-\gamma\tfrac{i}{2}\big)}
 e^{-ik^2/2-i\gamma\ln |k|+i\, \pi/4-\pi\gamma/4}, \\
 \xi_b={}& -\frac{\sqrt{\pi/\gamma}}
{\Gamma\big(\tfrac{1}{2}-\gamma\tfrac{i}{2}\big)}
 e^{-ik^2/2-i\gamma\ln |k|+ i\, 3\pi/4-\pi\gamma/4}.
\end{align}
\label{xisol}
\end{subequations}
Altogether, an arbitrary solution of \eqref{hamfulA} at $|k|\gg\delta$ is thus the linear combination
\begin{equation}
\psi(k)= e^{i\sigma_x\pi/4} \Psi(k)\bm{a}, \;\;
 \Psi(k) = \begin{pmatrix}
\eta_a & \eta_b\\ \xi_a & \xi_b
\end{pmatrix},\;\; \bm{a}= \begin{pmatrix}
a_1\\ a_2
\end{pmatrix},
\end{equation}
where $a_1,\, a_2$ are arbitrary coefficients. 

The approximate semiclassical solution of \eqref{hamfulA} reads \cite{Kaganov1983}
\begin{equation}
\psi_s(k)=\chi(k)\, e^{-i \int_{0}^k dk' k_z(k')+\phi_\mathrm{t}(k)},
\label{semiw}
\end{equation}
where $\chi(k)$ and $k_z(k)$ are determined by 
\begin{align}
&\big[\sigma_x\delta+ \sigma_y k + k_z^\pm(k)]\chi_\pm(k)= 0, \\
&\bigskip k^\pm_z(k)= \pm\sqrt{k^2+\delta^2}
\label{kz1}
\end{align}
and  $\phi_\mathrm{t}(k)$ is the topological phase shift accumulated on the orbit section between 
$k_y=0$ and $k_y=k$,
\begin{align}
\phi_\mathrm{t}(k) ={}& i \int_{0}^{k} dk_y'\,\chi^\dagger_\pm(k) \partial_k\chi_\pm(k)=\int_{0}^k dk'\frac{\delta}{2(\delta^2+(k')^2)}\nonumber \\
={}& \arctan(k/\delta)/2.
\end{align}
The first term in the exponent of the semiclassical wavefunction can be written as
\begin{subequations}
\begin{align}
\int_0^k dk' k_z^\pm(k') ={}& \pm\mathrm{sign}(k)f(k) 
+{\cal O}\big(\delta_0^2/k^2\big), \\
f(k) ={}& \tfrac{1}{2}\big[k^2+\delta_0^2
\big( \ln|2k/\delta_0|+1/2 \big)\big].
\end{align}
\label{phi0}
\end{subequations}

\begin{figure}
\includegraphics[scale=0.4]{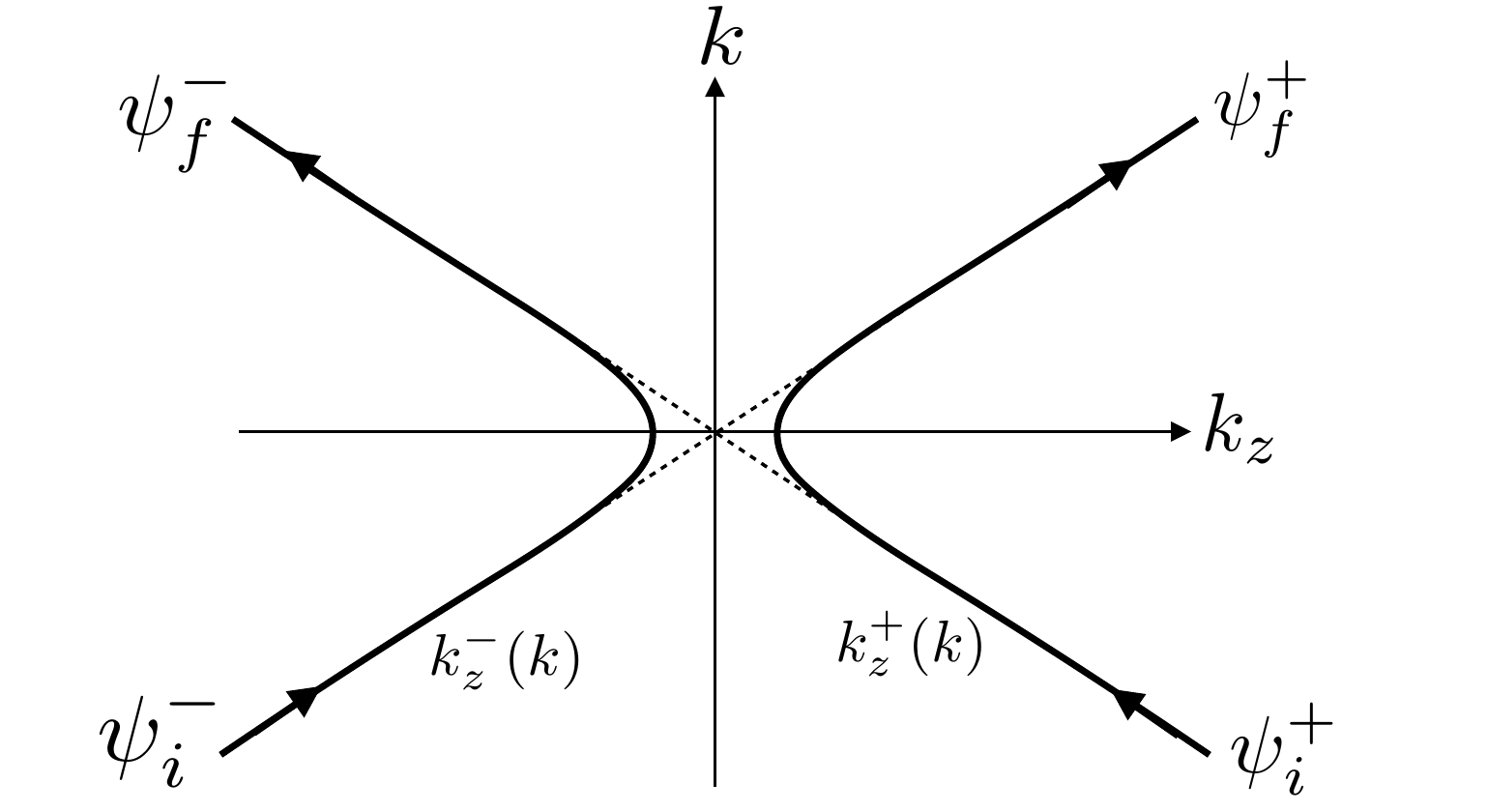}
\caption{Classical trajectories along the equi-energy contours $k_z^\pm(k)$ approaching and 
leaving the magnetic-breakdown region. The in- and outgoing scattering states, $\psi_i^\pm$ and 
$\psi_f^\pm$, respectively, are also indicated.}
\label{scattering}
\end{figure}

The basis for the scattering matrix is formed by the semiclassical wavefunctions at $k\ll-\delta$ as  
incoming states $\psi_i$ and at $k\gg\delta$ as  
outgoing states $\psi_f$, as indicated in Fig.\ \ref{scattering}. To leading order in $\delta/k$ we obtain
\begin{align}
&\psi_i^\pm =  e^{\pm if(k)}\frac{1}{\sqrt{2}}\begin{pmatrix}  \mp e^{i\pi/4}\\ e^{-i\pi/4} \end{pmatrix},\;\;\;\;\;\;\; \\
&\psi_f^\pm =  e^{\mp if(k)}\frac{1}{\sqrt{2}}\begin{pmatrix}  \mp  e^{-i\pi/4}\\ e^{i\pi/4}  \end{pmatrix}
\end{align}
and combine the scattering states  into matrices,
\begin{equation}
\Psi_i(k) = (\psi_i^+,\psi_i^-),\;\;\;\;\;\;\;\;\Psi_f(k) = (\psi_f^+,\psi_f^-).
\label{scatstates}
\end{equation}

We choose the coefficients of the full solution, $\bm{a}$, such that 
at $k_y\ll -\delta$ the full solution coincides with the incoming
state, $\Psi_i\bm{c}_i $, where according to \eqref{scatstates}, $\bm{c}_i=(1,0)$ corresponds to 
incoming state $\psi_i^+$ and $\bm{c}_i=(0,1)$ corresponds to 
incoming state $\psi_i^-$. At $k\gg \delta$ the phase and amplitude of the final states, combined in $\bm{c}_f$, 
is then determined by matching $\psi(k)$ with 
$\Psi_f\bm{c}_f$. Altogether, the matching conditions read
\begin{align}
&\Psi_i(k\ll -\delta)\bm{c}_i =  \Psi(k\ll -\delta)\bm{a},\\
&\Psi(k\gg \delta)\bm{a} = \Psi_f(k\gg\delta)\bm{c}_f.
\end{align}
Eliminating $\bm{a}$ we obtain the expression for the scattering
matrix $S$
\begin{equation}
\bm{c}_f = \underbrace{\Psi_f(k)^{-1}\Psi(k)\Psi^{-1}(-k)
\Psi_i(-k)}_{\equiv S} \bm{c}_i,\;\;\;\; k/\delta\to+\infty.
\end{equation}
Inserting the expressions $\Psi_f(k)$, $\Psi_i(k)$, and $\Psi(k)$ given above,  we obtain the scattering matrix \eqref{scatmattf} 
given in the main text.

\subsubsection{Topological Hamiltonian}

We consider the Schr\" odinger equation
\begin{align}
& \Big[\delta_n \sqrt{ u^2-1} \sin\theta \sigma_x+
 k \sqrt{ u^2-1}\sigma_y  + i\partial_k (u  + \sigma_z) + 
 \delta_n \cos\theta (1 + u \sigma_z)\Big]\psi=0.
\label{seIIA}
\end{align}
The semiclassical solution reads
\begin{equation}
\psi_s(k)=\chi(k)\, e^{-i \int_{0}^k dk' k_z(k')+\phi_\mathrm{t}(k)},
\label{semitop}
\end{equation}
where $\chi(k)$ and $k_z(k)$ are given by 
\begin{align}
& \Big[\delta_n \sqrt{ u^2-1} \sin\theta \sigma_x+
 k \sqrt{ u^2-1}\sigma_y +  k^\pm_z(k) (u  + \sigma_z) + 
 \delta_n \cos\theta (1 + u \sigma_z)\Big]\chi_\pm(k)=0, \label{hamtA}\\
&\bigskip k^\pm_z(k)= \pm\sqrt{k^2+\delta_n^2}.
\end{align}
The phase $\int_0^k dk' k_z^\pm(k') $  is in analogy 
to the non-topological case given by \eqref{phi0}
(with $\delta_0$ replaced by $\delta_n$). 
The topological phase shift is most easily obtained by considering the original Schr\" odinger equation 
\begin{align}
H_n'\psi ={}& \epsilon\psi, \\
H_n'={}& k_x^n\sigma_x+nk_x^{n-1}k_y\sigma_y+k_z\sigma_z+u k_z,
\end{align}
which is related to \eqref{seIIA} by a $k_z$ shift introduced in \eqref{kzshift} in the main text,
which leaves the phase shift accumulated between $k=0$ and $k=\pm\infty$ invariant.
The Hamiltonian is of the form of $H_1$. The topological 
phase shift thus calculates in analogy to the phase $\phi_\mathrm{t}^{1\pm}$
of the main text. Since by symmetry
the phase shift from $k=0$ to $k=\pm\infty$ is half the phase shift from $k=-\infty$ to $k=\infty$, we can use
Eq.\ \eqref{phibn1} to obtain
\begin{equation}
\phi_\mathrm{t}^{n\pm}(k=\infty)=\frac{\pi}{4}(1\mp\mathrm{sign}\theta)\pm\frac{\theta}{2},
\end{equation} \\
which is sufficient for the in- and outgoing states at $k=\pm\infty$.
Together with the spinors from \eqref{hamtA} the scattering states read
\begin{subequations}
\begin{align}
&\psi_i^\pm =  e^{\pm i f(k)-i\frac{\pi}{4}(1\mp\mathrm{sign}\theta)\mp i\theta/2}
\begin{pmatrix}  \mp i\sqrt{\frac{u-1}{2u}}\\ \sqrt{\frac{u+1}{2u}} \end{pmatrix},\\
&\psi_f^\pm =  e^{\mp if(k)+i\frac{\pi}{4}(1\mp\mathrm{sign}\theta)\pm i\theta/2}\begin{pmatrix} \pm i\sqrt{\frac{u-1}{2u}}\\ \sqrt{\frac{u+1}{2u}}  \end{pmatrix}.
\end{align}
\label{apsi}
\end{subequations}

To find the full solution $\psi(k)$ we multiply \eqref{seIIA} 
from the left with ${\cal M}=\mathrm{diag}[(u+1)^{-1},(u-1)^{-1}]$ and apply
a transformation given by 
\begin{equation}
T=-i\begin{pmatrix}
\frac{1-u}{\sqrt{u^2-1}} && \frac{u-1}{\sqrt{u^2-1}} \\ 1 && 1 
\end{pmatrix} \sigma_z e^{-i \sigma_y \theta/2},
\end{equation}
which leads to the differential equation of the Landau-Zener form \eqref{hamfulA},
\begin{align}
\hat{H} \hat{\psi}(k) &=\big( i\partial_k\sigma_0+\delta_n\, \sigma_x+ k\, \sigma_y \big)\hat{\psi}(k)= 0,
\end{align}
where $\hat{\psi}(k) = T^{-1}\psi(k)$ and $\hat{H} = T^{-1}\, {\cal M}\,\tilde{H} T$.
As in the non-topological case, we obtain the $S$ matrix by matching the full solution with the scattering states, 
$T^{-1}\psi^\pm_{i/f}$, which leads to the scattering matrix \eqref{scatmattf} with $\delta_0$ replaced 
by $\delta_n$. 

\section{Numerical results}
\label{numerics}

To give support to the analytical calculations, we numerically compute the offset $\gamma$ 
for type I and type II single, double, and triple Weyl nodes via numerical diagonalization of the Hamiltonian  
$\displaystyle{H'_n = H_n(\bm{k'}) + \eta {k'_z}^3 \sigma_z}$, 
with $H_n$ given by Eq.~\eqref{Hn} of the main text, and the regularizing term $\eta {k'_z}^3 \sigma_z$ to ensure closed
Fermi pockets in the case $u>1$, as discussed in the
main text. The magnetic field in the $x$ direction enters according to the Peierls substitution
$\bm{k'} = \bm{k} + \bm{A}$, with
\begin{equation}
k'_x = k_x,\;\; [k'_y, k'_z] = -iB\,.
\label{commutator}
\end{equation}
We make use of the ladder operators $a$ and $a^{\dagger}$ of the quantum oscillator
to construct momentum operators with the required properties.
Straightforwardly,  
\begin{equation}
k'_y = (a + a^{\dagger}) \sqrt{\frac{B}{2}},\;\;k'_z = i(a - a^{\dagger}) \sqrt{\frac{B}{2}}\,
\end{equation}
with $ [ a, a^{\dagger} ] = 1$, fulfil the commutator in Eq.~\eqref{commutator}.
\begin{figure}[t]
\begin{center}
 \subfigure[]{\includegraphics[width=0.7 \columnwidth]{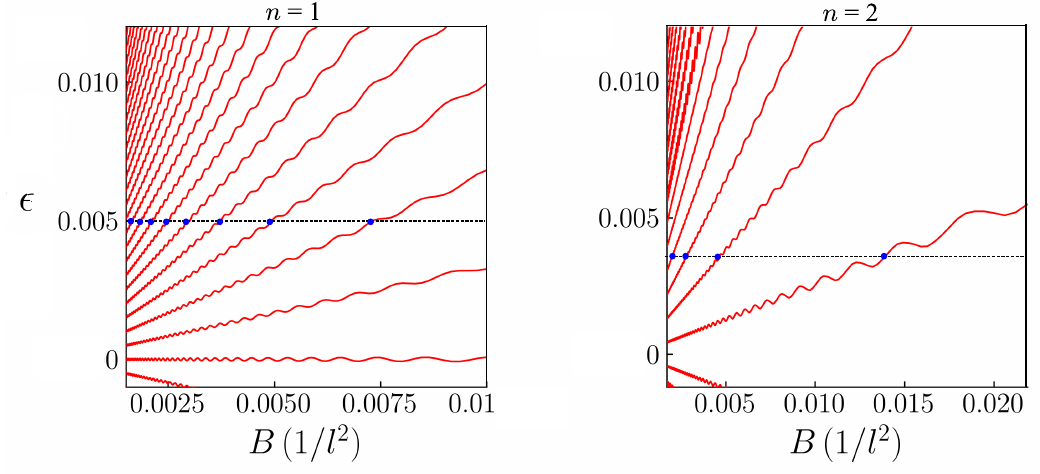}}
 \subfigure[]{\includegraphics[width=0.3\columnwidth]{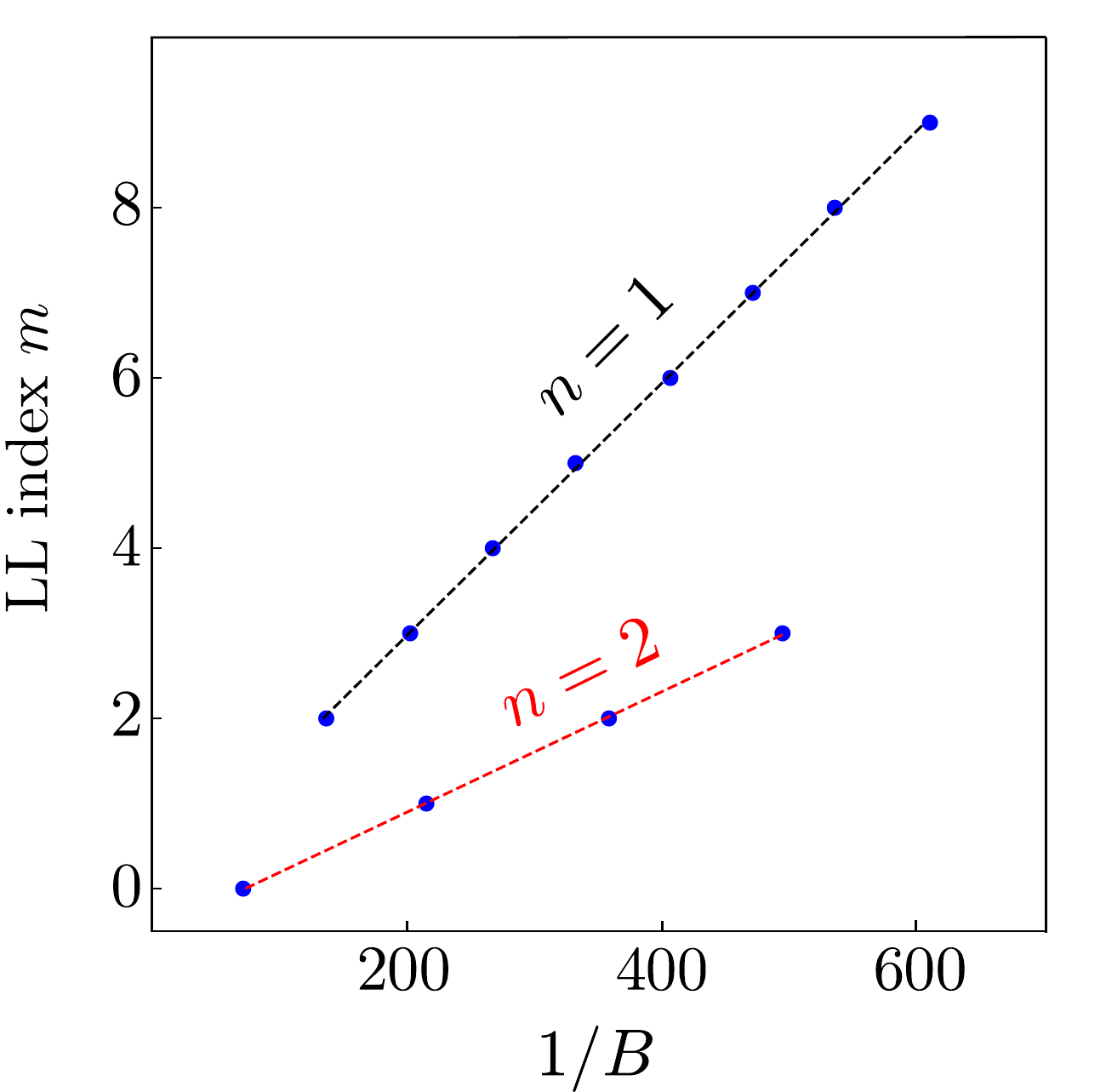}}
  \caption{(a) Landau fan diagrams for type II Weyl nodes with topological charge $n=1$ (left) and $n=2$ (rigth) at 
  $k_x=0.005$ and $k_x=0.03$, respectively.
  Other parameters are $u=1.6$, $\eta=0.1$, and $N_{cut}=2400$. (b) Landau-level (LL) index $m$ as a function of the 
  inverse field for $n=1,2$ at fixed energies indicated by the black dashed lines
  in (a). The dots correspond to the numerical data, the dashed lines to the linear fits according to Eq.~\eqref{ints}.
  \label{fansII}}
\end{center}
\end{figure}

\begin{figure}[t]
\centering
\includegraphics[width=0.4\columnwidth]{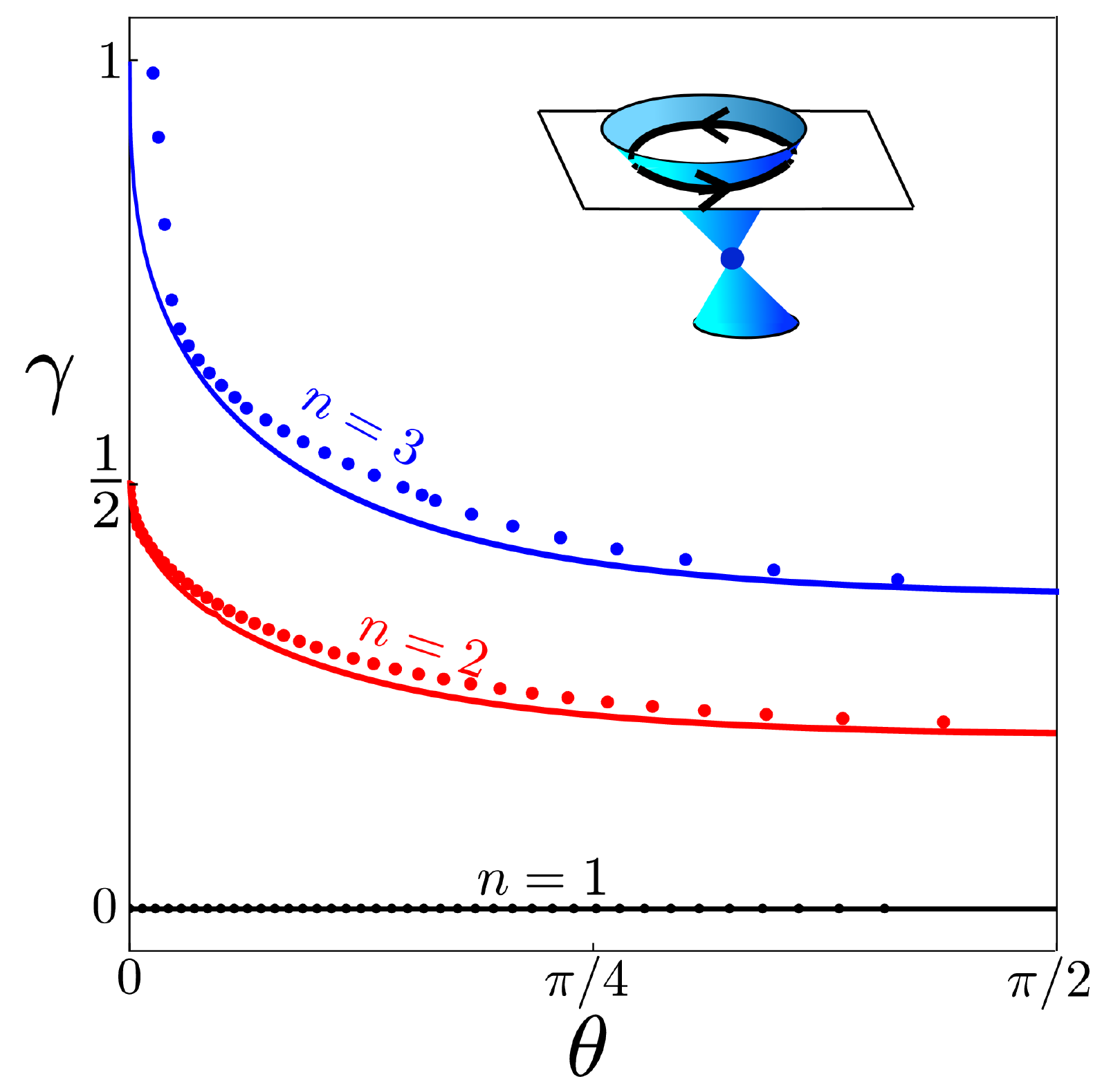}
\includegraphics[width=0.413\columnwidth]{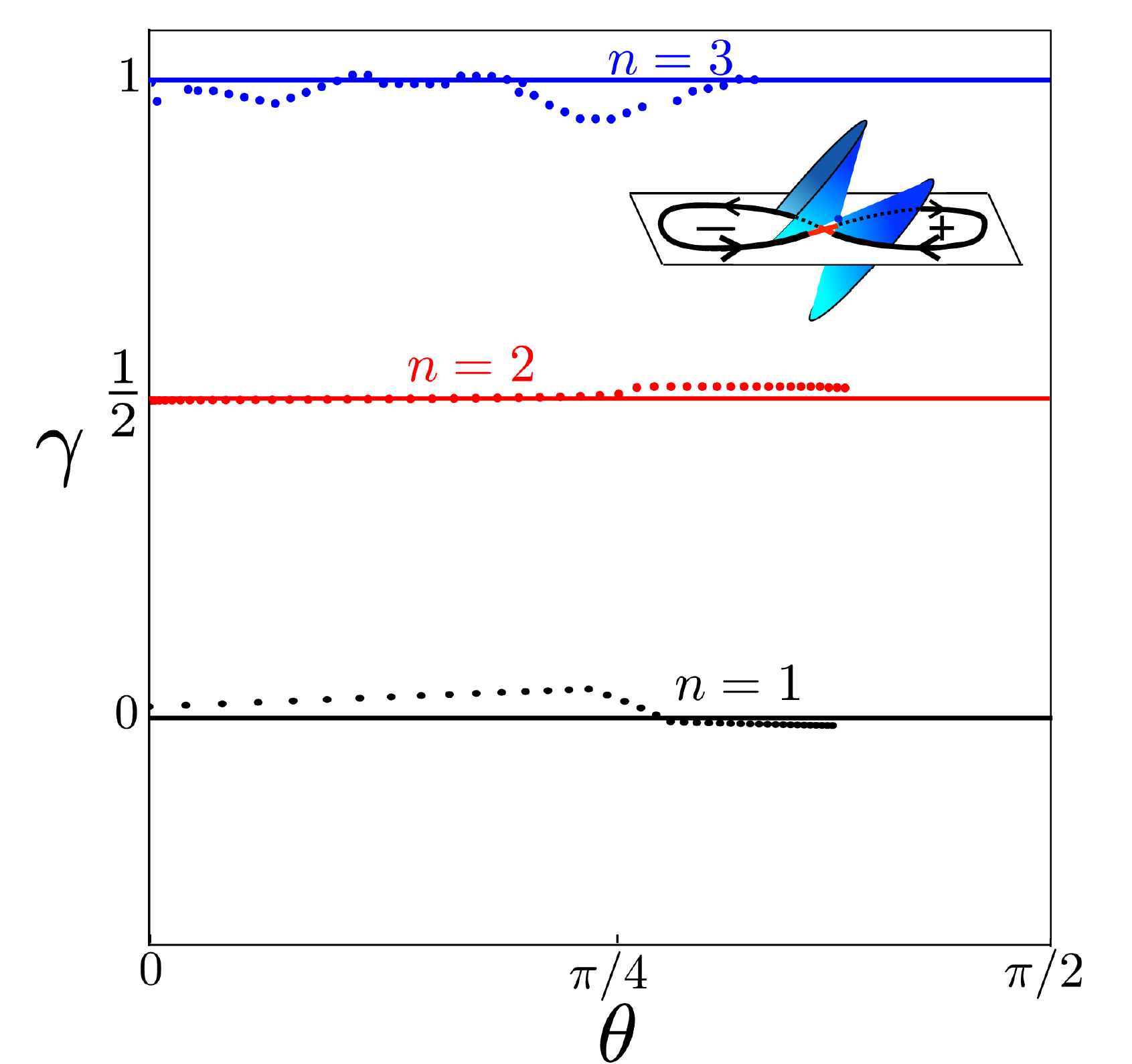}
\caption{Parameter dependence of the offset $\gamma$ (modulo one) for cyclotron orbits at type-I (left) and
type-II (right) Weyl nodes obtained numerically (dots),
compared to the analytical results (solid lines)  of the main text  [cf.\ Figs.\ \ref{phibI} and \ref{phibII}].
Parameters for numerical results are $u=1.6$ ($u=7$), $\eta=0.1$ ($\eta=1$), and $N_{cut}=2400$ ($N_{cut}=800$) for $n=1,2$ ($n=3$). The relatively large value of $u$ and $\eta$ for $n=3$ were necessary to access large values of $\theta$ [cf.\ Eq.\ \eqref{theta}], at the same time closing the contour at not too large momenta.}
\label{offsetI}
\end{figure}

The ladder operators are calculated in the basis of the Landau-level eigenstates (eigenstates 
of $a^{\dagger}a$), yielding the matrix elements
$ (a)_{ij} = \delta_{i,j+1}\sqrt{j} $ and $ (a^{\dagger})_{ij} = \delta_{i,j-1}\sqrt{i} $, respectively.
The lowest $l$ Landau levels are obtained by sparse diagonalization of the Hamiltonian
constructed from ladder operators truncated to $i,j \in[ 1 , N_{cut}]$ with $N_{cut}\gg l$, ensuring convergence of 
the eigenvalues with the value of $N_{cut}$. 

Fan diagrams, shown in Fig.~\ref{fansII}, are obtained by repeating this procedure at different values of the magnetic-field strength.
For type-II Weyl orbits the limit of unit breakthrough probability is never achieved in practice, 
resulting in oscillations on top of the fans \cite{Kaganov1983,Alexandradinata2017a}, 
 clearly visible in Fig.~\ref{fansII}, which however are not the subject of our present study.
In order to better extract the phase shift, we suppress these oscillations for $n=2$ and $n=3$ by 
averaging the energies over a range of magnetic fields containing
several oscillation peaks (dips).

At a fixed energy $\epsilon$, 
we extract the intercept fields  $\lbrace B_m \rbrace$, where $B_m$ is the value of the field
at which the energy of the Landau level with quantum number $m$ crosses $\epsilon$. 
The inverse of the intercept fields are then fitted to the quantization condition, Eq.~\eqref{qc} of the main text,
\begin{equation}
 \frac{1}{B_m} = \frac{2\pi}{S(\epsilon)} (m+\gamma)\,,
 \label{ints}
\end{equation}
where the zero-field area $S(\epsilon)$ 
enclosed by the equi-energy contour is calculated numerically from the dispersion at
$B=0$. The offset $\gamma$ modulo one is thus obtained as the only fitting parameter.

The results, shown in Fig.~\ref{offsetI}, are in good agreement with the analytical 
results presented in the main text for all the cases that we were able to address numerically.
The phase offset corresponding to the non-protected band-touching Hamiltonian (\ref{H0})
was numerically found to be $\gamma=1/2$ in a previous work by the authors~\cite{Bovenzi2017a}, also in agreement with 
analytics.

\end{document}